\documentclass[trackchanges, twocolumn]{aastex701}

\usepackage{amsmath}
\usepackage{multirow}
\usepackage{caption}
\graphicspath{{./}{}}

\begin{document}

\title{Einstein Probe Fast X-ray Transients Extend the Physical Parameter Space of Relativistic Jets}
%Highlight Tension with Standard Off-Axis GRB Models
%}

\author[orcid=0000-0001-7880-2407,gname='Connery',sname='Chen']{Connery Chen}
\affiliation{Nevada Center for Astrophysics, University of Nevada, 4505 S. Maryland Pkwy., Las Vegas, NV 89154-4002, USA}
\affiliation{Department of Physics and Astronomy, University of Nevada, 4505 S. Maryland Pkwy., Las Vegas, NV 89154-4002, USA}
\email[show]{connery.chen@unlv.edu}  

\author[orcid=0000-0002-8614-8721,gname=Yihan, sname='Wang']{Yihan Wang} 
\affiliation{Department of Astronomy, University of Wisconsin, Madison, WI 53706, USA.}
\email{wang3697@wisc.edu}

\author[orcid=0000-0002-9725-2524,gname=Bing, sname='Zhang']{Bing Zhang} 
\affiliation{The Hong Kong Institute for Astronomy and Astrophysics, the University of Hong Kong, Pokfulam Road, Hong Kong, China.}
\affiliation{Department of Physics, University of Hong Kong, Pokfulam Road, Hong Kong 999077, China.}
\email{bing.zhang@hku.edu}

\begin{abstract}
Fast X-ray Transients (FXTs) detected by the Einstein Probe (EP) mission possess exceptionally low spectral peak energies compared to typical long (Type II) GRBs. 
Some of these extragalactic transients show phenomenological similarities to X-ray flashes (XRFs), but the physical origins of FXTs remain uncertain.
In this work, we investigate EP-detected FXTs using various jet structures relevant to Type II GRBs and test the hypothesis that these FXTs belong to the intrinsically same population of known GRBs but viewed at large angles from the jet axis.
We apply detectability estimates to evaluate their distribution in the observed $E_{\rm  iso}$--$E_{\rm p}$ plane. We find that standard single- and multi-component jet structures can reproduce the energetics of off-axis events such as GRB 170817A and low-luminosity GRBs (llGRBs), while also yielding energetics consistent with XRFs at the lower end of the $E_{\rm iso}$--$E_{\rm p}$ continuum for moderately off-axis observers. However, viewing-angle effects of standard Type II GRBs alone cannot account for the $E_{\rm p}$ values observed in some energetic FXTs. This tension suggests that EP-detected FXTs are unlikely to be explained solely as classical GRBs viewed off-axis, and may instead probe relativistic explosions in a previously underexplored region of parameter space. In particular, these transients may be associated with lower Lorentz factors, reduced angular momentum in the collapsing core, or alternative jet structures and emission mechanisms. Our results motivate further studies to test these scenarios and constrain the physical properties of FXT progenitors and their outflows.
\end{abstract}

%% https://astrothesaurus.org

\section{Introduction} 
\label{sec:intro}

Fast X-ray transients (FXTs) are a growing class of luminous, short-duration X-ray events with spectral peaks within or below the soft X-ray band ($E_{\rm p} \leq 10\,\rm keV$). They share observational similarities with X-ray flashes (XRFs) discovered by early X-ray missions such as BeppoSAX and HETE-2, which were initially interpreted as either off-axis gamma-ray bursts (GRBs) or intrinsically softer explosions~\citep{sakamoto_04, sakamoto_05, irwin_16}. XRFs are generally considered the soft end of the long GRB population, characterized by low spectral peak energies and X-ray--dominated prompt emission~\citep[e.g.][]{zhang_04_quasi,zhang_04}.

Recent observations by the Einstein Probe (EP; \citealt{yuan_25_ep}) have significantly expanded the known sample of extragalactic FXTs, revealing bursts with soft spectra  ($E_{\rm p} \leq 10$ keV) and low isotropic energies. Some EP-detected events with $E_{\rm p} \geq 10$ keV are accompanied by gamma-ray emission detected by high-energy instruments~\citep{ep240315a, ep240801a, ep250404a}, while others show no detectable gamma-ray counterparts~\citep{ep240414a, ep250108a, dai_26}. In this work we refer to the former as ``EP-detected GRBs", and reserve the term ``FXTs" for EP-detected X-ray transients lacking detectable gamma-ray emission.

\setcounter{footnote}{0}
FXTs are commonly associated with broad-lined Type Ic (Ic-BL) supernovae at low redshift ($z \lesssim 1$), likely due to selection effects from their modest isotropic-equivalent luminosities, $L_{\rm iso}$~\citep{ep240414a, ep250108a}. However, more luminous FXTs have also been detected at higher redshifts, such as at $z = 1.53$~\citep{dai_26}. In contrast, EP-detected GRBs are observed at higher redshifts, with one event detected at $z=4.859$~\citep{liu_25_240315a}, and resemble canonical long (Type II\footnote{Throughout this work, we adopt the terminology introduced in \cite{zhang_06_type, zhang_07_type, zhang_09_type}: ``Type II'' GRBs refer to GRBs associated with the core collapse of massive stars~\citep{woosley_bloom_06_grb_sne}, and ``Type I'' GRBs refer to GRBs associated with compact-object mergers~\citep{eichler_89, narayan_92, berger_14}.}) GRBs with multiwavelength counterparts spanning gamma-rays to radio and associated with Ic-BL supernovae. A notable feature of EP-detected GRBs is that their soft X-ray emission both precedes and outlasts the gamma-ray emission.

While most FXTs are associated with Type II progenitors, recent observations suggest that the FXT population may be heterogeneous. Notably, a subset of EP-detected FXTs have been localized near passive or weakly star-forming galaxies, raising the possibility that some FXTs may originate from compact binary mergers (Type I GRBs) rather than massive-star collapse~\citep{becerra_26}. However, the evidence for compact-binary-merger associations remains circumstantial due to the absence of conclusive kilonova signatures and uncertainties in the host-galaxy associations.

These observational properties raise an important question: are FXTs simply canonical Type II GRBs observed off-axis, or do they represent a heterogeneous class of relativistic transients with multiple progenitor channels? Proposed interpretations for llGRBs and XRFs that may also be applicable to FXTs include baryon-loaded (``dirty'') fireballs~\citep[e.g.,][]{meszaros_91, pac_98, dermer_99,huang_02}, choked or failed jets~\citep[e.g.,][]{piran_17}, or a distinct progenitor class~\citep[e.g.,][]{irwin_16}.

In this work, we test whether FXTs can be explained as geometric variants of canonical Type II GRBs. We consider both single-component and multi-component structured jet models and explore their observable properties over a wide range of viewing angles. Using numerical simulations, we generate synthetic GRB populations and evaluate their consistency with observed samples.

We find that off-axis observers of structured jets can reproduce the energetics of llGRBs and XRFs. However, some FXTs occupy an even softer regime while maintaining relatively high luminosities, which cannot be explained by viewing-angle effects of canonical Type II GRBs alone. This suggests that at least some FXTs are unlikely to be simple off-axis GRBs and instead require alternative jet structures or additional emission physics.

In Section~\ref{sec:obs}, we present an observational dataset of GRBs and highlight several notable EP-detected GRBs and FXTs.  In Section~\ref{sec:model}, we summarize the numerical model and jet structures considered. Using our numerical model, in Section~\ref{sec:results} we perform data simulations consistent with Type II GRBs. We discuss the implications of our results in Section~\ref{sec:discussion} and conclude our findings in Section~\ref{sec:conclusion}.

\section{Observational Data}
\label{sec:obs}

GRB measurements with known redshifts are obtained from \cite{mp_20}. The sample includes 45 Type I and 275 Type II GRBs detected by various missions up to January 2019. The isotropic-equivalent energy $E_{\rm iso}$ ($1$--$10{,}000 \, \rm keV$) and intrinsic peak energy $E_{{\rm p},z}$ are provided for each burst. 
While we do not extend the observational sample beyond 2019, we include GRB 221009A~\citep{frederiks_23} and GRB 170817A~\citep{zhang_18} for comparison.

The number of FXTs without detectable gamma-ray counterparts remains small. In this work we include all published EP-detected FXTs (listed below in bold), as well as several EP-detected GRBs:

\begin{itemize}
    \item EP240315a~\citep{ep240315a}: A GRB jointly detected by Swift/BAT and Konus-Wind, with $E_{\rm p} = 283^{+65}_{-47}\,\rm keV$ and $E_{\rm iso} = 6.4^{+0.4}_{-0.8}\times10^{53}\,\rm erg$ ($1$--$10{,}000$ keV). The X-ray emission ($T_{90,X} = 1034 \pm 81\,\rm s$, $0.5$--$4\,\rm keV$) both precedes and outlasts the gamma-ray emission ($T_{90,\gamma} \sim 40\,\rm s$, $15$--$350\,\rm keV$ and $23$--$1618\,\rm keV$). The X-ray peak coincides with the hardest time-resolved spectrum. An optical counterpart yields $z = 4.859$.
    \item \textbf{EP240414a}~\citep{ep240414a}: An FXT with a very soft spectrum ($\beta \sim 3.1$), $E_{\rm p} \lesssim 1.3\,\rm keV$---likely peaking below the EP band---and $E_{\rm iso} = 5.3^{+0.8}_{-0.6}\times10^{49}\,\rm erg$ ($0.5$--$4\,\rm keV$). It is associated with the Type Ic-BL supernova SN2024gsa at $z=0.401$, located $26.3\pm0.1\,\rm kpc$ offset from a Seyfert I host galaxy.
    \item EP240801a~\citep{ep240801a}: An XRF with fluence ratio $S(25$--$50\,{\rm keV})/S(50$--$100\,{\rm keV}) = 1.67^{+0.74}_{-0.46}$, jointly detected by \textit{Fermi}/GBM. It has $E_{\rm p} = 14.90^{+7.08}_{-4.71}\,\rm keV$ and $E_{\rm iso} = 5.57^{+0.54}_{-0.50}\times10^{51}\,\rm erg$ ($1$--$10{,}000$ keV). X-ray emission ($T_{90,X} = 148.0\pm3.2\,\rm s$) precedes and outlasts the gamma rays ($T_{90,\gamma} = 22.30\pm9.92\,\rm s$). An optical counterpart was found at $z = 1.6734$.
    \item \textbf{EP250108a}~\citep{ep250108a}: An FXT with a very soft spectrum ($\beta \sim 4.2$), $E_{\rm p} \lesssim 1.8\,\rm keV$---likely peaking below the EP band---and $E_{\rm iso} \sim 10^{49}\,\rm erg$ ($0.5$--$4\,\rm keV$). No gamma-ray or radio emission is detected. It is associated with the Type Ic-BL SN2025kg at $z = 0.176$.
    \item EP250404a~\citep{ep250404a}: A GRB jointly detected by Fermi/GBM, with $E_{\rm p}=55.34^{+0.64}_{-0.51} \, \rm keV$ and $E_{\rm iso} = 3.05^{+0.01}_{-0.01}\times 10^{53} \, \rm erg$ ($1-10,000$ keV). The X-ray emission ($T_{90,X} \sim 300\,\rm s$) precedes and outlasts the gamma rays ($T_{90,\gamma} = 90.43^{+0.64}_{-0.37}\,\rm s$). An optical counterpart reveals $z = 1.88$.
    \item \textbf{EP241113a}~\citep{dai_26}: An FXT with peak energy $E_{\rm p} \leq 2.4\,\rm keV$ and $E_{\rm iso} \sim 1.4^{+0.7}_{-0.2} \times 10^{51}\,\rm erg$ ($0.5$--$4\,\rm keV$). No gamma-ray or radio emission is detected. An optical counterpart was identified at $z = 1.53$.
\end{itemize}

\begin{figure}
    \centering
    \includegraphics[width=\linewidth]{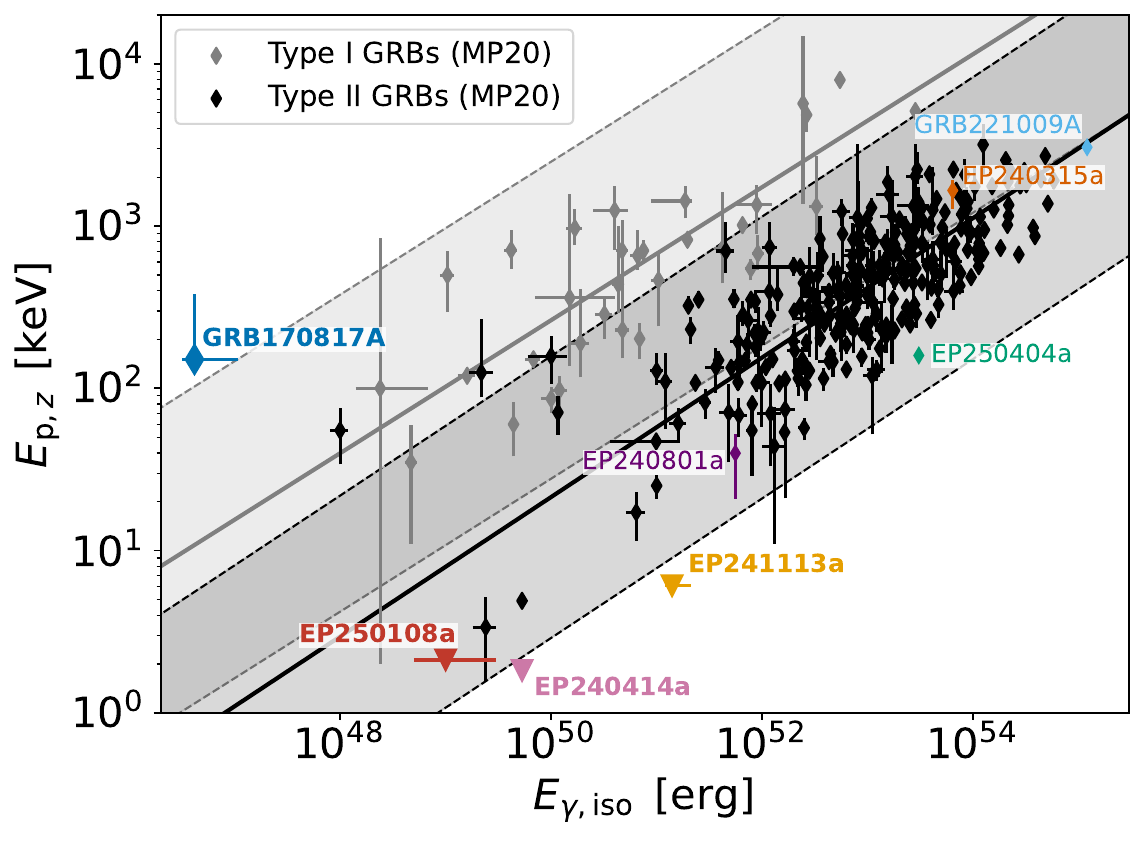}
    \caption{Observed GRBs with known redshift in the $E_{\gamma, \rm iso}$--$E_{{\rm p},z}$ plane~\citep{mp_20}. Grey and black diamonds represent Type I and Type II GRBs, respectively. The solid line and shaded region represent the Amati relation (Eq.~\ref{eq:amati}) and its $3\sigma$ scatter. Colored markers and corresponding labels indicate EP-detected FXTs and GRBs~\citep{liu_25_240315a, ep240414a, jiang_25_240801, ep250108a, ep250404a}, GRB 170817A~\citep{zhang_18} and GRB 221009A~\citep{frederiks_23}. EP-detected FXTs occupy a region of unusually low $E_{{\rm p},z}$ relative to $E_{\gamma, \rm iso}$, softer than the canonical Amati relation; their corresponding labels are bolded. Downward arrows represent upper limits.}
    \label{fig:obs_data}
\end{figure}

The observational dataset is shown in Fig.~\ref{fig:obs_data}, including both Type I and Type II GRBs with their respective Amati relations~\citep{mp_20}. The shaded regions correspond to the $3\sigma$ scatter of the observational data. The EP-detected GRBs and FXTs outlined above, including GRB 221009A~\citep{frederiks_23} and GRB 170817A~\citep{zhang_18} are denoted by special colors corresponding to their label in the lower-right list. Notice that EP-detected FXTs EP250108a, EP240414a and EP241113a pose upper limits very near or below the $3\sigma$ range of the Amati relation for Type II GRBs; EP-detected GRBs and XRFs, however, are consistent with the general population.

\section{Methods}
\label{sec:model}
\subsection{Prompt Emission Model}
To simulate observed GRBs and EP-detected FXTs, we adopt the numerical framework for prompt emission from relativistic jets introduced in \citet{chen_25_xray_grb}, which we summarize here.

The observed temporal profile of a GRB pulse is modeled with a fast-rise-exponential-decay (FRED) function \citep{norris_05}:
\begin{equation}
\label{eq:fred}
I(t) = \frac{A\lambda}{\exp \left(\tau_1/t + t / \tau_2 \right)},\quad t>0,
\end{equation}
where $A$ is a normalization constant, $\tau_1$ and $\tau_2$ are the rise and decay timescales, and $\lambda=\exp(2\mu)$ with $\mu = \sqrt{\tau_1/\tau_2}$. The pulse peak occurs at $t_{\rm peak} = \sqrt{\tau_1 \tau_2}$, the width measured between $1/e^3$ amplitudes is $\Delta t = \sqrt{9 + 12 \mu}$, and the pulse asymmetry is $\tau_2 / \Delta t$.

Although this functional form captures the general temporal morphology of long and short GRBs, it does not encode the full complexity of highly variable, multi-peaked light curves. Bursts with sharper or more intermittent spikes can yield systematically higher peak luminosity even for comparable total radiated energy. Our use of a FRED profile therefore provides a first-order estimate of the peak luminosity, appropriate for evaluating global population trends rather than detailed burst-level variability. 

The observed GRB spectrum is modeled with a Band function \citep{band_93}:
\begin{equation} 
\label{eq:band} 
N(E) = A \times \left\{ 
\begin{aligned} 
&E^{\alpha} \exp\left(\frac{E \, (\beta - \alpha)}{E_\text{b}}\right), & \text{for } E < E_{\text{b}}, \\ 
&E_{\text{b}}^{\alpha - \beta} \exp(\beta - \alpha) \, E^\beta, & \text{for } E \geq E_{\text{b}}, 
\end{aligned} 
\right. 
\end{equation}
where $A$ is the normalization, $\alpha$ and $\beta$ are the low- and high-energy photon indices, and $E_b = (\alpha - \beta) [E_{\rm p} / (2 + \alpha)]$ is the break energy, where $E_{\rm p}$ is the spectral peak.

The intrinsic gamma-ray energy per solid angle of the jet is
\begin{equation}
\epsilon_\gamma (\theta) \equiv \frac{dE_\gamma(\theta)}{d\Omega}.
\label{eq:eps}
\end{equation}
The initial Lorentz factor of the ejecta, $\Gamma_0$, is empirically related to the observed isotropic-equivalent energy $E_{\gamma, \rm iso}$ \citep{liang+10_relation, ghirlanda+11_relation}:
\begin{equation}
\label{eq:lg_relation}
\Gamma(E_{\gamma, \rm iso}) = \Gamma_0 \left(\frac{E_{\gamma, \rm iso}}{10^{52}\,\rm erg}\right)^{1/4},
\end{equation}
with $\Gamma_0 \sim 180$. Updated samples confirm a similar scaling~\citep{ghirlanda_18}.

The relativistic Doppler factor is
\begin{equation}
\label{doppf}
\mathcal{D} = \frac{1}{\Gamma(1-\beta\cos\alpha)},
\end{equation}
where $\alpha$ is the angle between the patch of emitting material and the observer's line of sight. For structured jets, we define the on- and off-axis Doppler factors, $\mathcal{D}_{\rm on}$ and $\mathcal{D}_{\rm off}$, and their ratio
\begin{equation}
\mathcal{R}_\mathcal{D} \equiv \frac{\mathcal{D}_{\rm off}}{\mathcal{D}_{\rm on}} < 1,
\end{equation}
which scales observed properties between observers:
\begin{align}
dt_{\rm off} &= \mathcal{R}_\mathcal{D}^{-1} \, dt_{\rm on}, \\
\epsilon_{\rm off} &= \mathcal{R}_\mathcal{D}^{3} \, \epsilon_{\rm on}.
\end{align}
The cubic dependence arises from relativistic beaming of isotropic emission in the comoving frame, where observed energy flux scales as $\mathcal{D}^3$ for a moving surface element~\citep{rl_79}.

The total energy per solid angle observed along an arbitrary line of sight $(\theta_v, \phi_v)$ is obtained by summing contributions from every jet patch at $(\theta_i, \phi_j)$:
\begin{equation}
\label{eq:eps_bar}
    \bar\epsilon_\gamma (\theta_0, \phi_0) = k_0 \frac{\sum_{i,j} \epsilon_{\gamma}(\theta_i, \phi_j) \mathcal{R}_\mathcal{D}^3 \Delta\Omega_{i,j}}{\sum_{i,j} \Delta\Omega_{i,j}},
\end{equation}
where $k_0$ is the normalization factor defined such that the on-axis isotropic-equivalent energy, $E_{\gamma, \rm iso} = 4\pi \bar\epsilon_\gamma(0,0)$, is equal to a given (observed) value.

At each $(\theta_i,\phi_j)$ patch, we adopt a phenomenologically motivated intrinsic on-grid spectrum $\mathcal{N}_{i,j}$ that is proportional to Band function (Eq.~\ref{eq:band}) with $E_{{\rm p},0}\sim 600$ keV, $\alpha=-1$, and $\beta=-2.3$~\citep{goldstein_16}. This prescription reproduces an overall Band-like spectrum when the emission from all patches is integrated~\citep{chen_25_xray_grb}. The peak energy scales with isotropic energy via the Amati relation \citep{amati_02}:
\begin{equation}
\label{eq:amati}
\log\left(\frac{E_{{\rm p},i}}{100\,{\rm keV}}\right) = a \log\left(\frac{E_{\rm iso}}{10^{51}\,{\rm erg}}\right) + b,
\end{equation}
with updated coefficients for Type I ($a_{\rm I} = 0.41 \pm 0.05$ and $b_{\rm I}  = 0.83 \pm 0.06$) and Type II ($a_{\rm II}  = 0.43 \pm 0.03$ and $b_{\rm II}  = -0.24 \pm 0.06$) GRBs~\citep{mp_20}. The Amati relation is used here as a phenomenological prescription to connect spectral peak energy and total isotropic energy in the prompt emission. Our goal is not to test the origin of the correlation itself, but to ensure that the simulated GRB population reproduces the observed distribution of prompt emission properties. We note that applying the Amati relation to off-axis emission is an extrapolation, since the empirical correlation is derived from the observed population of predominantly (near) on-axis events. This empirical relation is most reliable near the jet core and becomes increasingly uncertain at large viewing angles. We therefore treat the simulated population boundary as indicative rather than precise, and focus on gross deviations rather than marginal ones.

The total gamma-ray energy per patch is
\begin{equation}
\label{eq:e_tot}
\epsilon_{\gamma,i,j} = A_{i,j} \int_{E_1}^{E_2} E \, \mathcal{N}_{i,j}(E) , dE,
\end{equation}
where $\mathcal{N}_{i,j}(E)$ is the observer-frame spectral shape per solid angle for the $(i,j)$ patch (e.g., a Band function; Eq.~\ref{eq:band}), and $A_{i,j}$ is the normalization constant chosen such that the integrated spectrum reproduces the prescribed gamma-ray energy per solid angle, $\epsilon_{\gamma,i,j}$, of that patch.

% The observed luminosity light curve per patch, $\mathcal{L}_{i,j} = \partial L_{i,j}/\partial \Omega$, satisfies
% \begin{equation}
% \epsilon_{\gamma,i,j} = \int_{t_1}^{t_2} \mathcal{L}_{i,j}(t) , dt,
% \end{equation}
% with the temporal profile prescribed by a FRED function (Eq.~\ref{eq:fred}).

The observed spectra of off-axis patches are Doppler-shifted:
\begin{equation}
\label{eq:spec_obs}
\mathcal{N}_{i,j,\rm obs}(E) = \mathcal{N}_{i,j}(E) \mathcal{R}_\mathcal{D}(\theta_0, \phi_0, \theta_i, \phi_j),
\end{equation}
and summed across all patches to obtain the total observed spectrum per solid angle:
\begin{equation}
\label{eq:N_E_tot}
\mathcal{N}_{\rm tot}(E) = \frac{\sum_{i,j} \mathcal{N}_{i,j,\rm obs}(E) \Delta \Omega_{i,j}}{\sum_{i,j} \Delta \Omega_{i,j}}.
\end{equation}
The total observed energy per solid angle and isotropic-equivalent energy in the detector band follows as
\begin{align}
\bar\epsilon &= \int_{E_1}^{E_2} E \, \mathcal{N}_{\rm tot}(E) \, dE,\\
E_{\rm iso} &= 4 \pi \bar\epsilon.
\label{eq:e_iso}
\end{align}

For an observer located at $(\theta_{\rm obs},\phi_{\rm obs})$, the emission from each emitting surface element $(\theta_i,\phi_j)$ is transformed from the source frame to the observer frame. We denote the emission time in the source (engine) frame by $t_{\rm eng}$ and the photon arrival time measured by a distant observer by $t_{\rm obs}$.

The emitting radius of each patch is given by
\begin{equation}
    R_{{\rm em},i,j}(t)
    =
    \frac{\beta_{i,j} c\, t_{{\rm eng},i,j}}
    {1-\beta_{i,j}},
\end{equation}
where $\beta_{i,j} = \sqrt{1 - \Gamma_{i,j}^{-2}}$. For a finite-duration pulse in a structured outflow, the emitting radius evolves with the source-frame emission time.

The corresponding observer-frame photon arrival time is
\begin{equation}
\label{eq:eats_obs}
    t_{{\rm obs},i,j}
    =
    t_{{\rm eng},i,j}
    +
    \frac{R_{{\rm em},i,j}}{c}
    \left(1-\cos\theta_{\rm w}\right),
\end{equation}
where $\theta_{\rm w}$ is the angular separation between the $(\theta_i,\phi_j)$ patch and the observer's line of sight. This treatment naturally accounts for equal-arrival-time surface (EATS) effects arising from both angular and radial propagation delays across the outflow.

We denote by $\mathcal{L}_{i,j}$ the source-frame luminosity per unit solid angle of the $(i,j)$ patch. The corresponding observed luminosity is Doppler-transformed as
\begin{equation}
    \mathcal{L}_{i,j,\rm obs}
    =
    \mathcal{L}_{i,j}\,
    \mathcal{R}_{\mathcal D}^{4}(\theta_{\rm obs},\phi_{\rm obs},\theta_i,\phi_j).
\end{equation}

The observed emission from each surface element is mapped onto a common observer-time grid, $t_{\rm obs}$ using Eq.~\ref{eq:eats_obs}, thereby accounting for EATS-induced temporal broadening across the structured outflow. The total observed luminosity per unit solid angle is thus obtained by summing over all emitting surface elements:
\begin{equation}
\label{lc_obs}
    \mathcal{L}_{\rm tot}(t_{\rm obs})
    =
    \frac{
    \sum_{i,j}
    \mathcal{L}_{i,j,\rm obs}(t_{\rm obs})
    \,\Delta\Omega_{i,j}
    }
    {
    \sum_{i,j}
    \Delta\Omega_{i,j}
    }.
\end{equation}

The isotropic-equivalent luminosity light curve is then
\begin{equation}
    \label{eq:L_iso}
    L_{\rm iso}(t_{\rm obs}) = 4\pi \mathcal{L}_{\rm tot}(t_{\rm obs})
\end{equation}

\subsection{Jet Structure}
\label{sec:jet_struct}

The angular structure of GRB jets is commonly parameterized using one (or a combination) of the following functional forms that describe how the gamma-ray energy per unit solid angle, $\epsilon_\gamma(\theta)$, varies with polar angle, $\theta$:

I. Tophat,
\begin{equation}
\epsilon_\gamma(\theta) = \epsilon_{\gamma,\,0} \times 
\begin{cases}
1, & \theta \leq \theta_{\rm j}, \\
0, & \theta > \theta_{\rm j},
\end{cases}
\label{eq:tophat}
\end{equation}
where $\epsilon_{\gamma, \, 0}$ is the peak energy per solid angle in gamma-rays and $\theta_j$ is the jet half opening angle.

II. Gaussian~\citep{zhang_02},
\begin{equation}
\epsilon_\gamma(\theta) = \epsilon_{\gamma,\,0} \times
\begin{cases}
  \exp\left[-\theta^2 / ( 2 \theta_j^2 ) \right], & \theta \leq \theta_{\rm cut}, \\
0, & \theta > \theta_{\rm cut},
\end{cases}
\label{eq:gaussian}
\end{equation}
where $\theta_{\rm cut}$ is the cutoff angle. 

III. Power-law~\citep{rossi_02},
\begin{equation}
\epsilon_\gamma(\theta) = \epsilon_{\gamma,\,0} \times 
\begin{cases}
1, & \theta \leq \theta_j, \\
\left( \dfrac{\theta}{\theta_j} \right)^{-k}, & \theta_j < \theta \leq \theta_{\rm cut}, \\
0, & \theta > \theta_{\rm cut},
\end{cases}
\label{eq:powerlaw}
\end{equation}
where $k$ is the power-law decay index. 

The cutoff angle, $\theta_{\rm cut}$, is introduced to limit the angular extent of energy within a structured jet, but its exact value is not well constrained observationally. Analyses of Type II GRBs suggest that efficient $\gamma$-ray emission is restricted to a narrow region around the jet core, with little flux contributed from wide wings \citep{ben_nakar_19}. In contrast, numerical modeling of the prompt and afterglow emission from the Type I GRB 170817A~\citep[e.g.,][]{ryan_24, chen_25_xray_grb} suggests a broader angular structure ($\theta_{\rm cut}/\theta_{j} \sim 7$). These different angular extents in energy between Type I and Type II GRBs likely reflect intrinsic differences in their circumburst environments: jets from compact binary mergers propagate through relatively low-mass ejecta, permitting broader but sparser wings, whereas jets in massive stars traverse denser stellar envelopes, which can yield more collimated cores with sharply truncated emission at wider angles. We emphasize that while the concept of a wing or cutoff angle is useful for structured-jet models, its precise value remains uncertain and can vary substantially between bursts.

The intrinsic structure of GRB jets is also significantly shaped by the central engine and jet breakout timescales. 
For Type I GRBs arising from NS mergers, a short jet-launching delay timescale allows the relativistic jet to propagate through the compact dynamical ejecta with minimal obstruction, maintaining a narrow, highly relativistic core that may gradually develop a Gaussian-like profile through interaction with the ambient medium~\citep{geng_19}. In contrast, delayed launches encounter an extended merger ejecta, causing the jet to deposit significant energy laterally into the ejecta, naturally producing a multi-component structure consisting of a narrow ultra-relativistic core surrounded by slower, broader wings.
For Type II GRBs originating from the core collapse of massive stars, the jet must always first penetrate the dense, extended stellar envelope. If the central engine remains active for a duration $t_{\rm eng}$ longer than the jet breakout timescale, $t_{\rm breakout}$, then the jet can successfully break out of the stellar material and produce a GRB with duration $t_{\rm GRB}\sim t_{\rm eng} - t_{\rm breakout}$. If $t_{\rm eng} < t_{\rm breakout}$, the jet is choked and most of its energy will be deposited into the stellar envelope, producing predominantly cocoon emission~\citep{bromberg_11}.

These processes naturally yield a range of angular energy distributions, from sharply collimated cores to multi-component structures. To capture this diversity, we consider both single- and multi-component jet structures in our following analysis.

\subsubsection{Single-component Jet Structure}
\label{sec:single-comp}

A single-component jet structure provides a simple model for describing relativistic outflows in GRBs: the angular distribution of energy and Lorentz factor follows a single functional form (e.g., tophat, Gaussian, or power-law).
Here, we adopt a Gaussian energy profile (Eq.~\ref{eq:gaussian})~\citep{zhang_02}, which yields a smooth decline away from the jet axis and naturally accommodates both bright on-axis GRBs and fainter off-axis events~\citep{lamb_kobayashi_17}. The Lorentz factor profile is modeled using Eq.~\ref{eq:lg_relation}. Hydrodynamical simulations support the existence of angular structure within GRB jets~\citep{gottlieb_18_170817}, and GRB 170817A was successfully interpreted by off-axis emission from a Gaussian jet~\citep{gg_18_170817, ryan_etal_20_170817, lk_18_gaussian, troja_18_struct_jet, chen_25_xray_grb}. 

A single-component jet can be defined by the following set of parameters:
\begin{itemize}
    \item $E_\gamma$: The total beaming-corrected energy;
    \item $\theta_j$: The jet opening-angle, or characteristic Gaussian width;
    \item $\theta_{\rm cut}$: The cutoff angle, also referred to as $\theta_{\rm wing}$ in some contexts.
    \item $a_{\rm I}$, $b_{\rm I}$ ($a_{\rm II}$, $b_{\rm II}$): Amati relation parameters for Type I (II) GRBs, per Eq.~\ref{eq:amati}~\citep{mp_20}
    \item $\tau_1$, $\tau_2$: FRED pulse parameters, per Eq.~\ref{eq:fred}~\citep{norris_05}
\end{itemize}
For a given set of parameters defining a jet, one can place an observer at an arbitrary viewing angle $\theta_v$ and compute the observed $E_{\gamma, \rm iso}$ (Eq.~\ref{eq:e_iso}) and $L_{\gamma, \rm iso}$ (Eq.~\ref{eq:L_iso}) according to the model presented in Section~\ref{sec:model}.

\subsubsection{Multi-component Jet Structure}
\label{sec:multi-comp}

A more flexible alternative is a multi-component jet structure, in which a narrow, tophat jet core is surrounded by a power-law wing formed by the jet--ejecta interaction:
\begin{equation}
\epsilon(\theta) = \epsilon_{\gamma, 0} \times\begin{cases}
1, & \theta \le \theta_{\rm j}, \\
A_{\rm w} \left(\frac{\theta}{\theta_{\rm j}}\right)^{-k_{\rm w}}, & \theta_j < \theta \leq \theta_{\rm cut}, \\
0, & \theta > \theta_{\rm cut},
\end{cases}
\end{equation}
where $\theta_{\rm j}$ is the half opening angle of the jet core, $A_{\rm w}$ is the wing normalization factor, $k_{\rm w}$ is the power-law index, and $\theta_{\rm cut}$ is the cutoff angle. The Lorentz factor profile is modeled using Eq.~\ref{eq:lg_relation}.
The total beaming-corrected energy of the outflow is normalized to $E_\gamma$. Such a jet structure has been introduced to interpret the broad-band data of the brightest-of-all-time GRB 221009A \citep[e.g.][]{zhang_24,zheng_24}. 

Physically, the wing component represents shock-heated ejecta and material entrained during jet propagation through the circumburst environment which produces broader emission regions surrounding the ultra-relativistic core. The parameters $A_{\rm w}$ and $k_{\rm w}$ control the relative energy content and angular profile of the extended wing structure, respectively, thereby regulating the contribution of wide-angle emission to off-axis observers. For sufficiently large viewing angles, emission from the wing is expected to dominate the observed signal. 

\subsection{Model Parameters}
\label{sec:sim}

\begin{figure*}
    \centering
    \includegraphics[width=.9\linewidth]{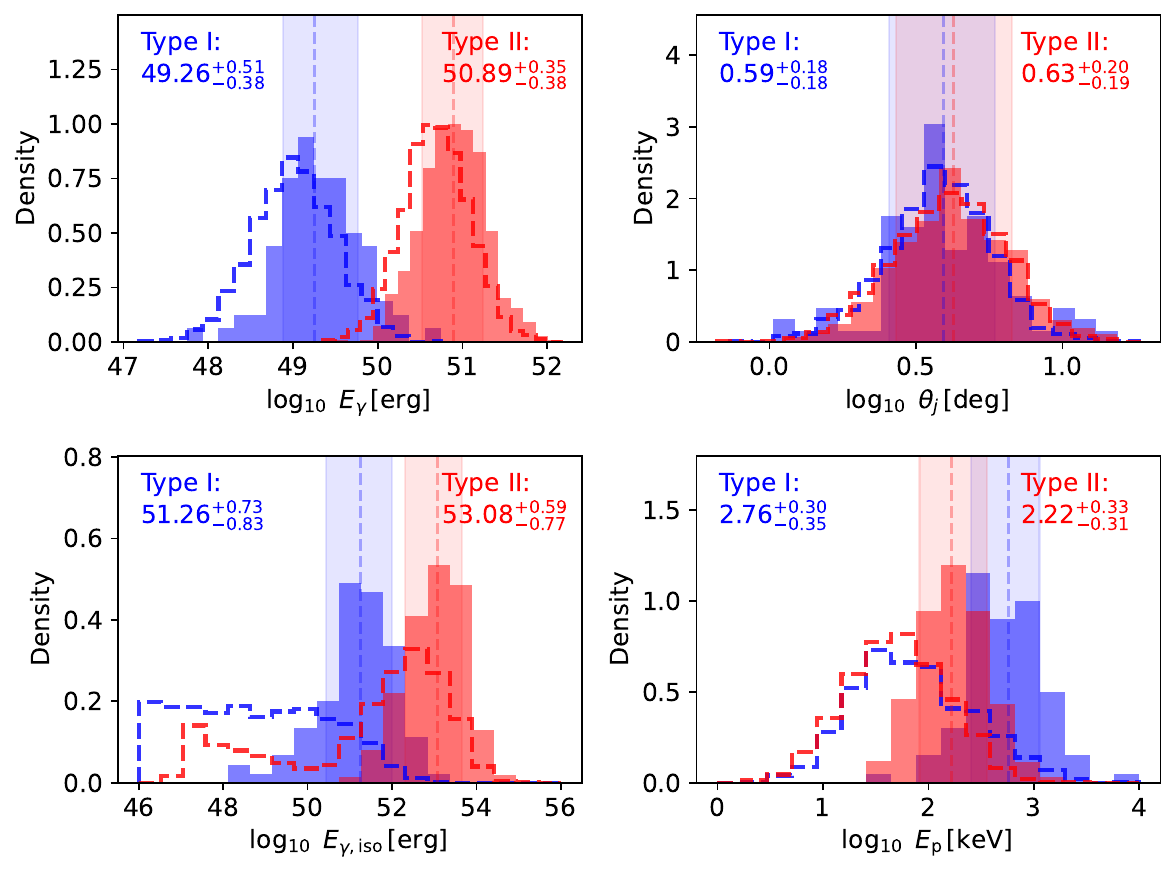}
    \caption{
    Single-component model parameter distribution densities for simulated events. Type I GRBs are shown in blue and Type II GRBs are shown in red. The filled histogram represents Swift/BAT-detectable simulated events (see Section~\ref{sec:detectability}), while the dashed step histogram shows the full simulated population, including events below the Swift/BAT detection threshold.}
    \label{fig:single_dist}
\end{figure*}

\begin{figure*}
    \centering
    \includegraphics[width=.9\linewidth]{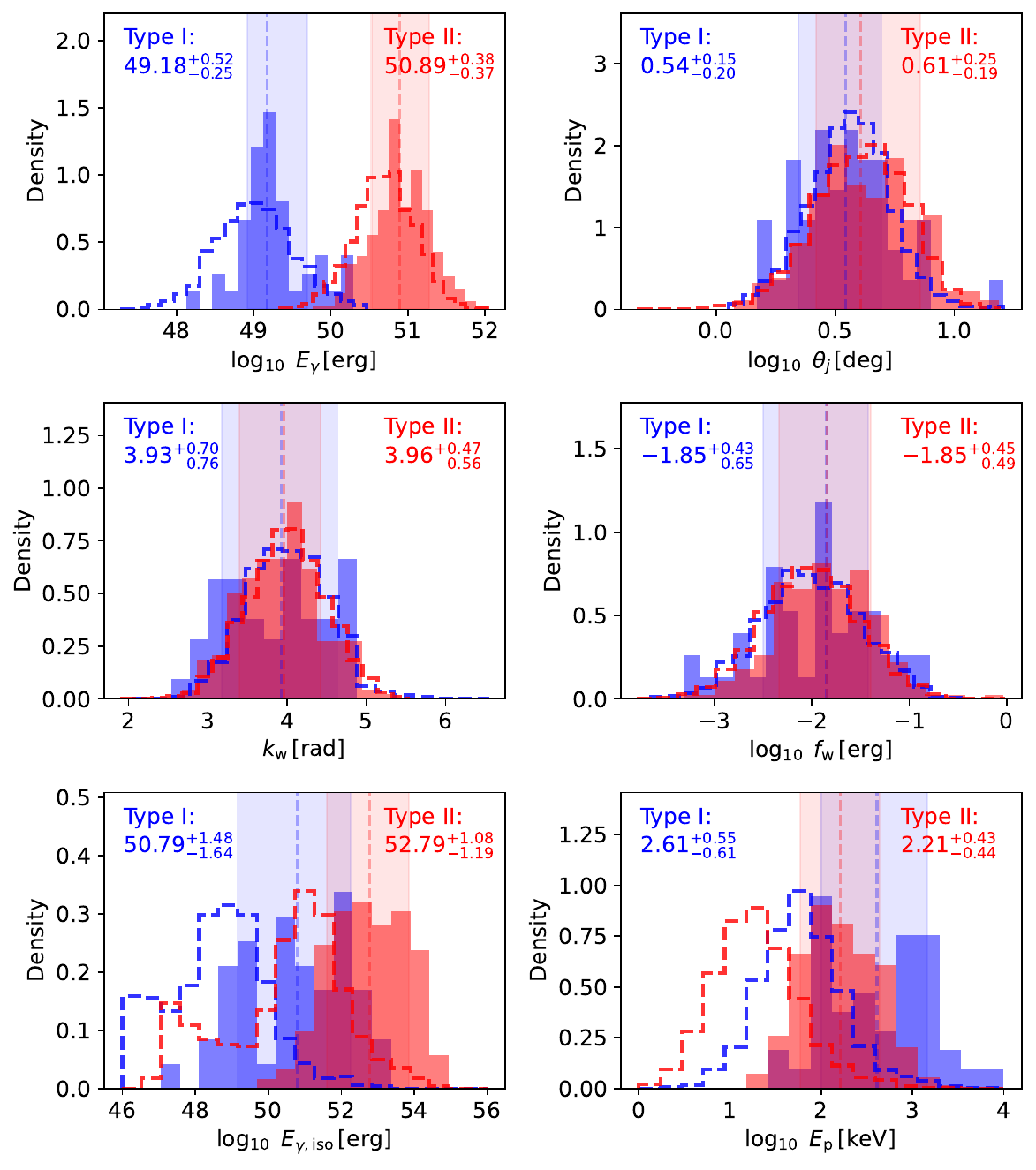}
    \caption{Same as Fig.~\ref{fig:single_dist} but for the multi-component model.}
    \label{fig:multi_dist}
\end{figure*}

We simulate Type I and Type II GRBs with the single-component (Section~\ref{sec:single-comp}) and multi-component (Section~\ref{sec:multi-comp}) jet structures. The input parameter distributions are given in Tables~\ref{tab:single_dist} and~\ref{tab:multi_dist}, and the corresponding sampled distributions are shown as dashed step histograms in Fig.~\ref{fig:single_dist} and~\ref{fig:multi_dist}.

\begin{table}
\centering
\caption{Parameter distributions for single-component Type I and Type II GRB models. $\mu$ is the median, always represented in linear space, and $\sigma$ is the standard deviation in linear space (for normal distributions) or $\log_{10}$ space (for log-normal distributions). For $\theta_v$, $\sigma$ denotes the allowed range, with sampling uniform in solid angle ($\propto \sin\theta_v$). $a_{\rm I}$ ($b_{\rm I}$) and $a_{\rm II}$ ($b_{\rm II}$) represent the Amati relation parameters for Type I and Type II GRBs, respectively.}
\begin{tabular}{l | c c | c c | c}
\hline
\multirow{2}{*}{\textbf{$\mathcal{P}$}} 
    & \multicolumn{2}{c|}{\textbf{Type I}} 
    & \multicolumn{2}{c|}{\textbf{Type II}} 
    & \multirow{2}{*}{\textbf{Distribution}} \\
\cline{2-3} \cline{4-5}
    & $\mu$ & $\sigma$ & $\mu$ & $\sigma$ &  \\
\hline
$E_\gamma \, [\textrm{erg}]$ 
    &$10^{49}$ & $0.5$ 
    &$10^{50.7}$ & $0.5$ 
    & log-normal \\

$\theta_j \, [\textrm{deg}]$ 
    & $3$ & $0.2$ 
    & $3$ & $0.2$ 
    & log-normal \\

$\theta_{\rm cut} \, [\textrm{deg}]$ 
    & $6\theta_j$ & ---
    & $4\theta_j$ & ---
    & fixed \\

$\theta_v \, [\textrm{deg}]$ 
    & --- & $0$--$30$ 
    & --- & $0$--$30$ 
    & $\sin$-uniform \\

$a_{\rm I} \, (a_{\rm II})$ 
    & $0.41$ & $0.05$ 
    & $0.43$ & $0.03$ 
    & normal \\

$b_{\rm I} \, (b_{\rm II})$ 
    & $0.83$ & $0.22$ 
    & $-0.24$ & $0.18$ 
    & normal \\

$\tau_1$ 
    & $0.004$ & $0.4$ 
    & $0.4$ & $0.5$
    & log-normal \\

$\tau_2$ 
    & $0.04$ & $0.4$ 
    & $4$ & $0.5$
    & log-normal

\end{tabular}
\label{tab:single_dist}
\end{table}

\begin{table}
\centering
\caption{Parameter distributions for multi-component Type I and Type II GRB models. In addition to the parameters defined in Table~\ref{tab:single_dist}, the wing component introduces two additional parameters: the normalization at the jet--wing boundary, $A_w$, and power-law index, $k_w$.}
\begin{tabular}{l | c c | c c | c}
\hline
\multirow{2}{*}{\textbf{$\mathcal{P}$}} 
    & \multicolumn{2}{c|}{\textbf{Type I}} 
    & \multicolumn{2}{c|}{\textbf{Type II}} 
    & \multirow{2}{*}{\textbf{Distribution}} \\
\cline{2-3} \cline{4-5}
    & $\mu$ & $\sigma$ & $\mu$ & $\sigma$ &  \\
\hline
$E_\gamma \, [\textrm{erg}]$ 
    & $10^{49}$ & $0.5$ 
    & $10^{50.7}$ & $0.5$ 
    & log-normal \\

$\theta_j \, [\textrm{deg}]$ 
    & $3$ & $0.2$
    & $3$ & $0.2$
    & log-normal \\

$k_w$ 
    & 4 & $0.5$
    & 4 & $0.5$
    & normal \\

$A_w$ 
    & $10^{-2}$ & $0.5$
    & $10^{-2}$ & $0.5$
    & $\log$-normal \\

$\theta_{\rm cut} \, [\textrm{deg}]$ 
    & $6\theta_j$ & ---
    & $4\theta_j$ & ---
    & fixed \\

$\theta_v \, [\textrm{deg}]$ 
    & --- & $0$--$30$
    & --- & $0$--$30$
    & $\sin$-uniform \\

$a_{\rm I} \, (a_{\rm II})$ 
    & $0.41$ & $0.05$
    & $0.43$ & $0.03$
    & normal \\

$b_{\rm I} \, (b_{\rm II})$ 
    & $0.83$ & $0.22$
    & $-0.24$ & $0.18$
    & normal \\

$\tau_1$ 
    & $0.008$ & $0.4$ 
    & $0.01$ & $0.5$
    & log-normal \\

$\tau_2$ 
    & $0.05$ & $0.4$ 
    & $0.6$ & $0.5$
    & log-normal

\end{tabular}
\label{tab:multi_dist}
\end{table}

These parameter distributions are chosen to be broadly consistent with observational constraints~\citep{fong_etal_15_grb1_parameters, escorial_etal_23_grb1_parameters, goldstein_16}, while maintaining the quasi-universal jet structure~\citep{zhang_04_quasi} assumption, \textbf{in which intrinsic variations in the jet opening angle are modest and viewing angle remains the primary driver of the observed diversity in $E_{\rm iso}$. Since the viewing-angle effect is primarily governed by the ratio $\theta_v/\theta_j$, and the distribution of $\theta_j$ is relatively narrow (0.2 dex) while $\theta_v$ spans a much wider range, the resulting variation in $\theta_v/\theta_j$ (and consequently in $E_{\rm iso}$) is dominated by changes in viewing angle. This trend is evident in Figs.~\ref{fig:sim_type2}, \ref{fig:sim_type1}, and \ref{fig:sim_KW}, where the viewing-angle color coding of the simulated data shows that larger viewing angles generally correspond to lower values of $E_{\rm iso}$.}

For a particular GRB type, the total jet energy $E_\gamma$ is kept equal in both the single- and multi-component jet structures. The geometric parameters $\theta_j$ and $\theta_{\rm cut}$ are also kept equal in both cases, but their interpretations differ: in the single-component jet structure, $\theta_j$ sets the characteristic width of the Gaussian profile, whereas in the multi-component jet structure it defines the half opening angle of the tophat core. 
Beyond $\theta_j$, the single-component jet structure continuously follows the smooth Gaussian profile while the multi-component jet structure transitions to the power-law wing.
The full jet structure is truncated at $\theta_{\rm cut}$, set to be a factor of several $\theta_j$, and is the same in both prescriptions.

While the goal of this work is to test whether FXTs can be explained purely as geometric variants of canonical Type II GRBs, we also simulate Type I GRBs for parameter validation. 
Although observational and theoretical studies suggest that Type II GRBs are typically more energetic and more narrowly collimated than Type I GRBs~\citep[e.g.,][]{frail_01, fong_etal_15_short_grb, zhang_04, bromberg_18}, we adopt a simplified approach in which both populations share the same characteristic angular scale, $\theta_j$. This choice allows us to isolate the effects of total energy and outer angular structure on the observed properties, without introducing additional variability from the jet core width. We emphasize that this assumption does not reflect the full physical diversity of GRB jets, but instead enables a controlled comparison between the two populations.
Accordingly, in our models, Type I GRBs are assigned lower total energy $E_\gamma$ and larger cutoff angle $\theta_{\rm cut}$, as shown in Tables~\ref{tab:single_dist} and~\ref{tab:multi_dist}.

%For a particular jet structure, Type I and Type II GRBs are assigned the same characteristic angular scale $\theta_j$, but different total energy $E_\gamma$ and cutoff angle $\theta_{\rm cut}$. 
%Holding $\theta_j$ constant across both populations eliminates the core width as a source of variability, allowing us to isolate the effects of total energy and outer angular structure on the observed properties. 
%Observationally, Type II GRBs are generally inferred to be more energetic and narrowly collimated than Type I GRBs, based on jet break measurements and afterglow modeling~\citep[e.g.,][]{frail_01, fong_etal_15_short_grb}. Numerical simulations of collapsar jets relevant to Type II GRBs similarly indicate strong collimation due to interaction with the stellar envelope, producing narrow, energetic outflows, whereas compact-object merger jets relevant to Type I GRBs tend to exhibit wider angular structures~\citep[e.g.,][]{zhang_04, bromberg_18}.

\subsection{Detectability}
\label{sec:detectability}

In this work, we adopt the Swift Burst Alert Telescope (BAT) as the primary reference for detectability. BAT provides one of the largest and most uniform samples of GRBs with well-characterized trigger properties, making it a convenient baseline for comparing simulated populations with observations. The detectable simulated events of all figures in the main text are based on Swift/BAT detectability.

For Swift/BAT, the $5\sigma$ detection threshold is parameterized as~\citep{baumgartner_13}
\begin{equation}
    f_{{\rm BAT},5\sigma} = 2.832 \times 10^{-8} \left(\frac{T}{1 \, {\rm  s}} \right) ^{-1/2} \rm erg \, cm^{-2} s^{-1},
\end{equation}
where $T$ is the exposure time over which the photon flux is integrated.

\begin{figure}
    \centering
    \includegraphics[width=\linewidth]{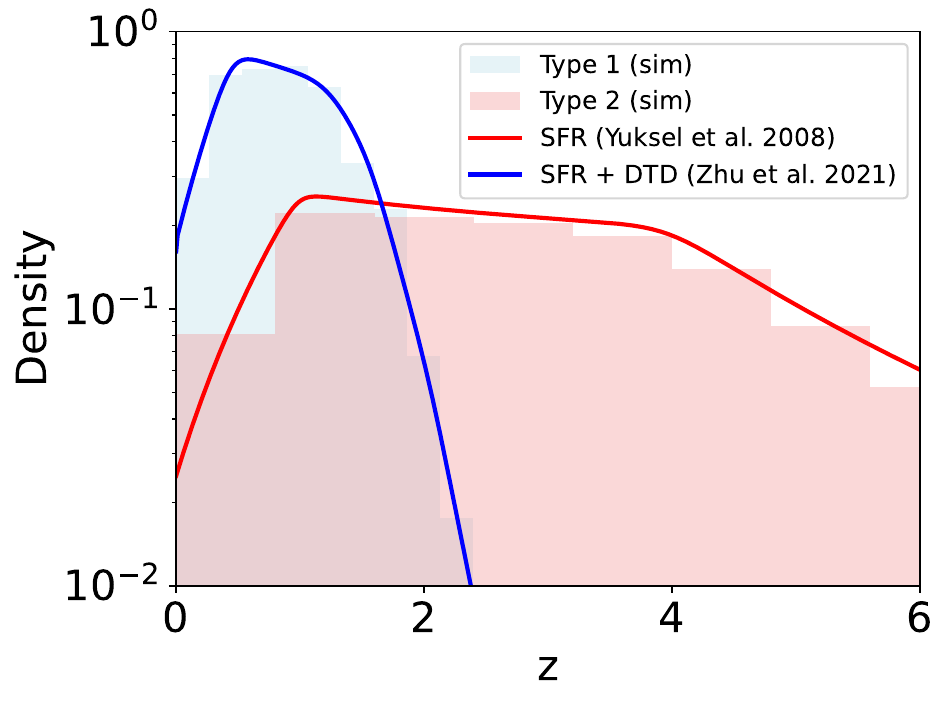}
    \caption{Intrinsic redshift distribution of Type I (blue) and Type II (red) GRBs. The simulated data are plotted as histograms, and model distributions are plotted as solid lines. 
    The intrinsic redshift distribution is model-independent and representative of both single- and multi-component jet structures.}
    \label{fig:z_dist}
\end{figure}

We assign each simulated burst a redshift drawn from population models appropriate to its progenitor class. Type II GRBs are associated with the core collapse of massive stars and are expected to trace the cosmic star-formation rate (SFR)~\cite{yuksel_etal_08_sfr}:
\begin{equation}
\mathrm{SFR}(z) = R_0 \left[ 
(1+z)^{a \eta} + \left(\frac{1+z}{B}\right)^{b \eta} + \left(\frac{1+z}{C}\right)^{c \eta} 
\right]^{1/\eta},
\end{equation} 
where $R_0 = 0.02 \, M_\odot \rm \, yr^{-1} \, Mpc^{-3}$ is the normalization factor, $a = 3.4$, $b = -0.3$, and $c = -3.5$ are the power-law slopes of the three segments, and $\eta=-10$ controls the sharpness of the transitions. \textbf{The breaks occur at $z_1=1$ and $z_2=4$, with the corresponding normalization constants $B = (1 + z_1)^{1-a/b} \simeq 5000$ and $C = (1 + z_1)^{(b-a)/c}(1 + z_2)^{1-b/c} \simeq 9$.
}

Type I GRBs, originating from compact binary mergers, follow a SFR-delayed distribution determined by the merger timescale~\citep[e.g., a log-normal delay-time distribution;][]{wanderman_piran_15}, which may be parameterized as~\citep{zhu_etal_21_redshift_dist}:
\begin{equation}
\begin{aligned}
R_{\rm I}(z)
=
\Biggl[
&(1+z)^{4.131\eta}
\\
&
+
\left(\frac{1+z}{22.37}\right)^{-0.5789\eta}
+
\left(\frac{1+z}{2.978}\right)^{-4.735\eta}
\\
&+
\left(\frac{1+z}{2.749}\right)^{-10.77\eta}
+
\left(\frac{1+z}{2.867}\right)^{-17.51\eta}
\\
&+
\left(\frac{1+z}{3.04}\right)^{
-\left(
0.08148
+
\frac{z^{0.574}}{0.08682}
\right)\eta
}
\Biggr]^{1/\eta},
\end{aligned}
\end{equation}
where $\eta = -5.51$. 
The resulting redshift distributions of the simulated sample are shown in Fig.~\ref{fig:z_dist}.

The observed flux in the Swift/BAT band can be estimated as 
\begin{equation}
F_{\rm BAT} = \frac{L_{\rm BAT}}{4 \pi d_L^2},
\end{equation}
where $L_{\rm BAT}$ is the observed luminosity in the Swift/BAT energy range (15--150 keV)---obtained following the method described in Section~\ref{sec:model}---and $d_L$ is the luminosity distance to the source. 

To evaluate detectability, we slide several integration windows of duration $T$ across the flux light curve, spanning $64 \, \rm ms \le T \le 64 \, s$, and compute the average flux in each window, $\langle F_T \rangle$. Each $\langle F_T \rangle$ is then compared to the BAT $5\sigma$ detection threshold, $f_{{\rm BAT},5\sigma} = 2.832 \times 10^{-8} (T / 1 \, {\rm s})^{-0.5} \rm \, erg \, cm^{-2} \, s^{-1}$. If $\langle F_T \rangle > f_{{\rm BAT},5\sigma}$ in any window (for example, in the $1 \, \rm s$ window if $\langle F_{1 \, \rm s} \rangle > 2.832 \times 10^{-8} \rm \, erg \, cm^{-2} \, s^{-1}$) then the burst is marked detectable, and its duration is quantified by the $t_{90}$ interval: the time during which $5\%$ to $95\%$ of the total detected fluence is accumulated. 

We also approximate Konus-Wind detectability in Appendix~\ref{app:kw_detectability} for the same sample of simulated Type II GRBs and find that the median $E_{\rm p}$ of KW-detectable bursts is $\sim 280 \, \rm keV$, compared to $\sim 160 \, \rm keV$ for the Swift/BAT-detectable sample, reflecting the higher-energy bandpass of KW.

We simulate $1{,}500$ Type I and $7{,}000$ Type II GRBs according to the parameter distributions in Tables~\ref{tab:single_dist} and \ref{tab:multi_dist}, yielding approximately $45$ and $275$ Swift/BAT-detectable events, respectively. 
In Figs.~\ref{fig:single_dist} and~\ref{fig:multi_dist}, filled histograms represent Swift/BAT detectable bursts, while dashed step histograms show the full simulated population, including undetectable bursts, for Type I and Type II GRBs, respectively. 
Detectable bursts are preferentially drawn from the high-$E_\gamma$ tail, and wider jets are marginally over-represented because they remain observable over a broader range of viewing angles, partially offsetting their lower on-axis brightness. 
%The impact of viewing angle is evident in the $E_{\gamma, \rm iso}$ distribution, where off-axis observers measure substantially lower energies, many of which fall below the detection threshold.
Detector selection effects are apparent in the observed $E_{\rm p}$ distribution, where the peak energy of detected bursts largely depends on the detector bandpass.

\section{Results}
\label{sec:results}
\begin{figure*}
    \centering
    \includegraphics[width=.9\linewidth]{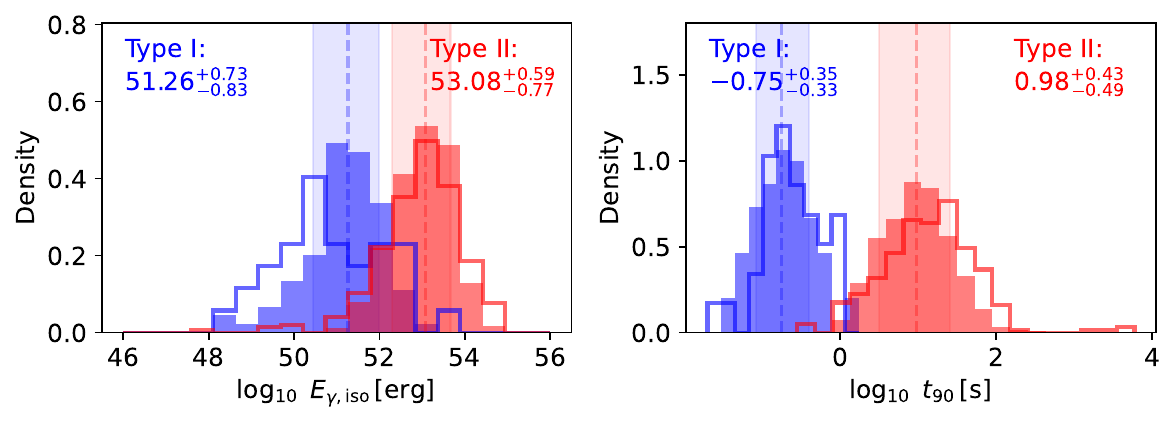}
    \includegraphics[width=.9\linewidth]{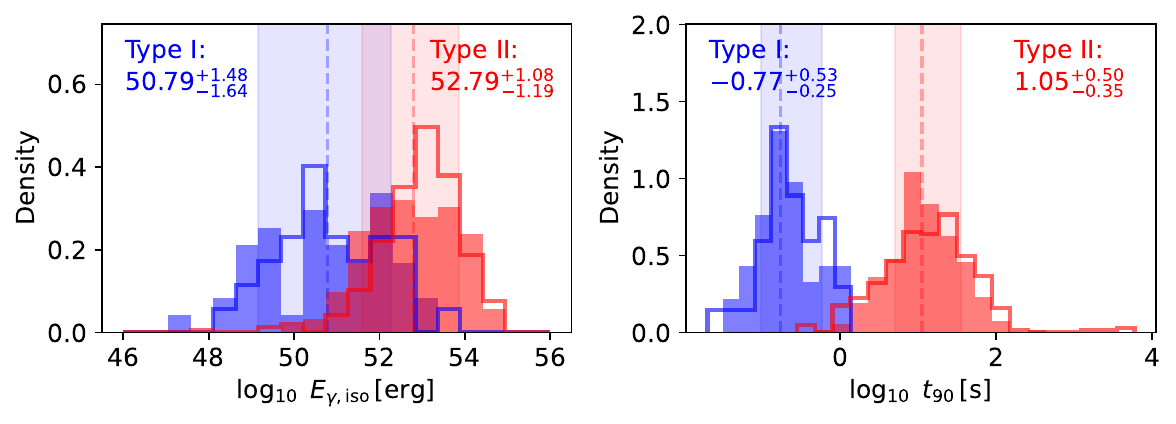}
    \caption{
    $E_{\gamma,\rm iso}$ and $t_{90}$ distribution densities of simulated and observed Type I (blue) and Type II (red) GRBs. The upper panels correspond to the single-component jet structure, and the lower panels correspond to the multi-component jet structure. Simulated events are represented by filled histograms, and the observational sample from \cite{mp_20} is shown as solid step histograms. The median and $1\sigma$ range of the distributions are indicated by dashed vertical lines and shaded regions, respectively, with values listed at the top of each panel. 
    }
    \label{fig:dist_E_iso_t90}
\end{figure*}

\begin{figure*}
    \centering
    \includegraphics[width=.49\linewidth]{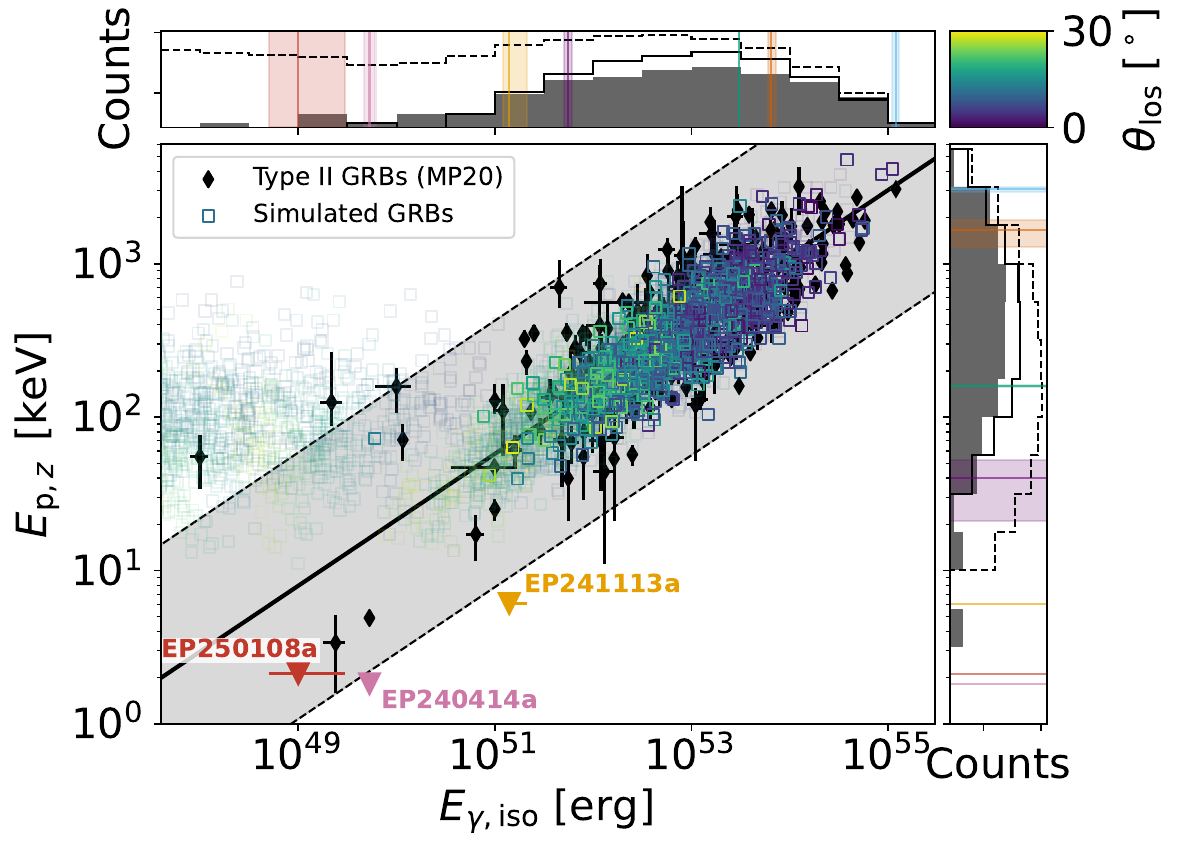}
    \includegraphics[width=0.49\linewidth]{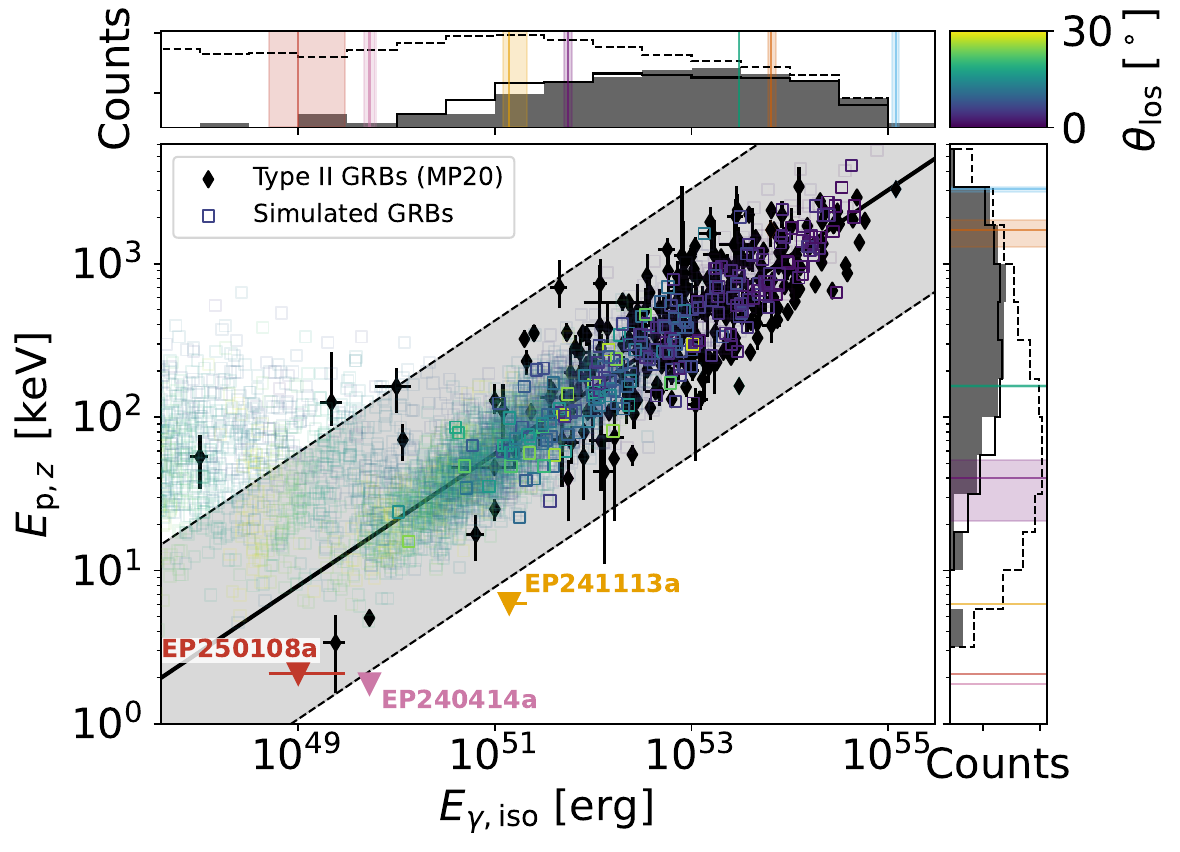}
    \caption{Simulated and observed Type II GRBs on the $E_{\gamma, \rm iso}$--$E_{{\rm p}, z}$ plane. Simulated bursts are shown for single-component (left panel) and multi-component (right panel) jet structures, with parameters sampled according to Tables~\ref{tab:single_dist} and~\ref{tab:multi_dist}, respectively. Simulated bursts are color-coded by viewing angle $\theta_v$, with opacity indicating Swift/BAT detectability (opaque for detectable bursts, semi-transparent for non-detections). 275 observed Type II GRBs (black diamonds) from~\cite{mp_20} are plotted alongside the Amati relation and its 3$\sigma$ scatter. FXTs (Section~\ref{sec:obs}) are plotted in select colors with corresponding labels in the lower-right list. Marginal histograms along the top and right axes show the distributions of $E_{\gamma, \rm iso}$ and $E_{{\rm p}, z}$, respectively, where solid lines represent the detectable (full) sample and the filled histogram represents the observational sample.}
    \label{fig:sim_type2}
\end{figure*}

\begin{figure*}
    \centering
    \includegraphics[width=.49\linewidth]{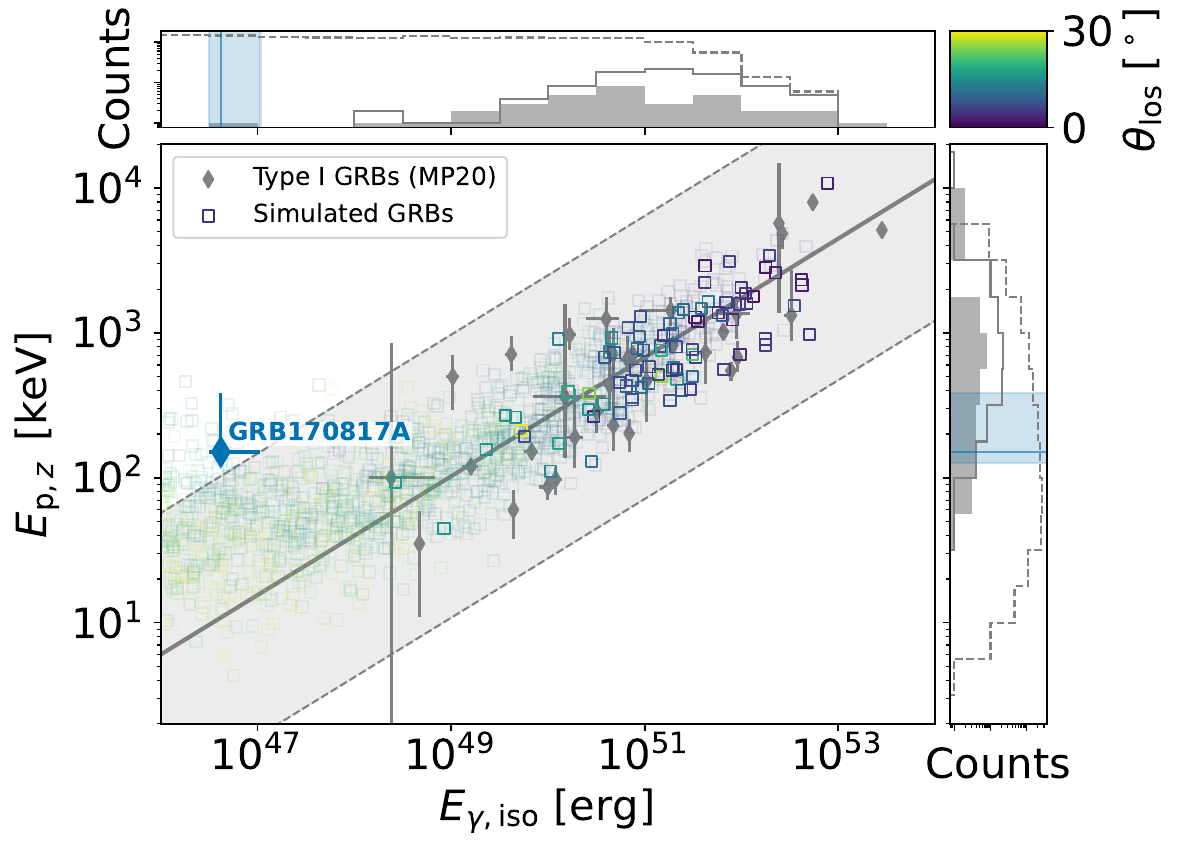}
    \includegraphics[width=0.49\linewidth]{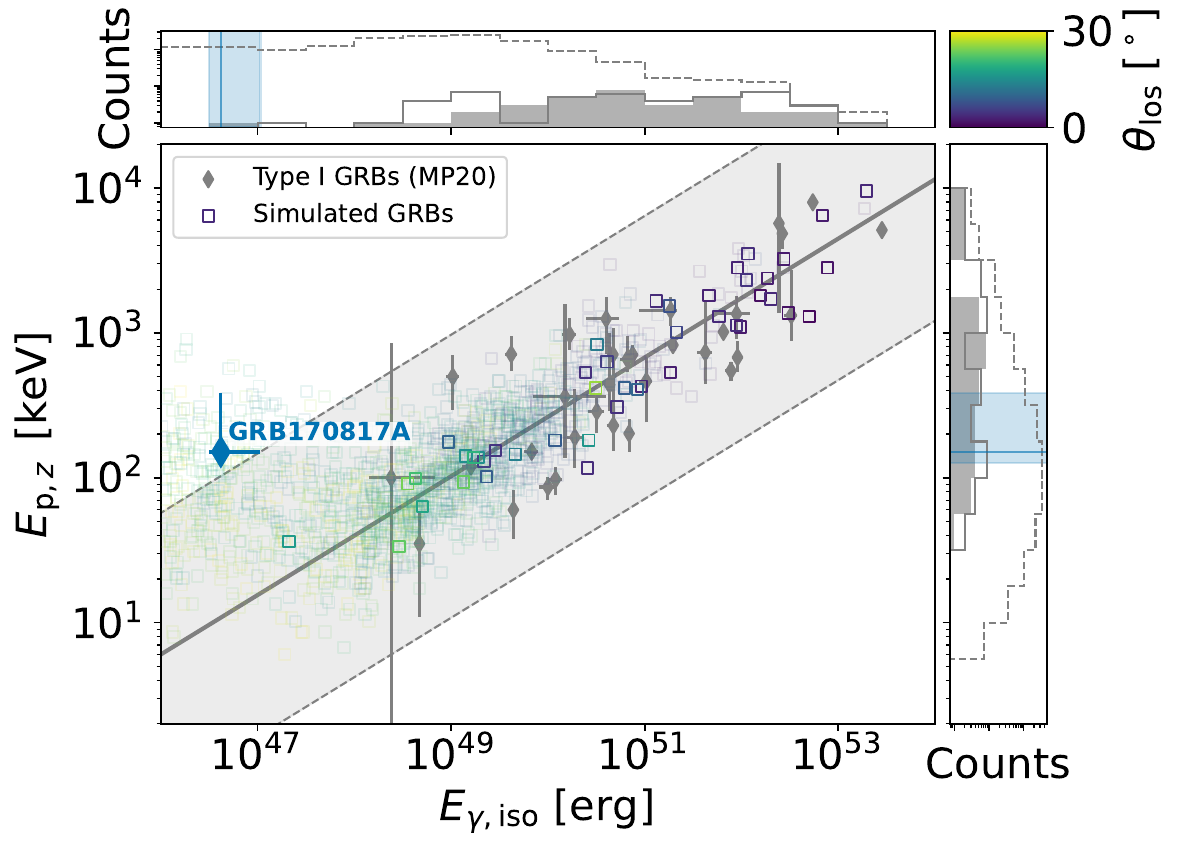}
     \caption{Same as Fig.~\ref{fig:sim_type2} but for Type I GRBs. 45 observed Type I GRBs (gray diamonds) are obtained from~\cite{mp_20}.
     GRB 170817 is included for completeness.}
    \label{fig:sim_type1}
\end{figure*}

The observed distribution densities of $E_{\gamma, \rm iso}$ and $t_{90}$ for the simulated bursts are shown in Fig.~\ref{fig:dist_E_iso_t90}, with the upper panels corresponding to the single-component model and the lower panels to the multi-component model.
Swift/BAT-detectable bursts are represented by filled histograms, while the observational sample is shown by the step histogram. 
The $E_{\gamma, \rm iso}$ distribution density is narrowly concentrated in the single-component model, but more broadly spread in the multi-component model due to the extended wing that distributes substantial energy to large angles.
The $t_{90}$ distribution primarily influences detectability, and its measurement and implications are discussed in Section~\ref{sec:detectability}. 
The overall consistency between the detectable simulated events and the observational sample serves to justify our choice of parameter distributions for Type I and Type II GRBs.

Figures~\ref{fig:sim_type2} and~\ref{fig:sim_type1} present the simulated and observed GRB populations in the $E_{\gamma, \rm iso}$--$E_{{\rm p}, z}$ plane. 
Across both single- and multi-component jet structures, viewing-angle effects introduce a systematic trend where off-axis bursts deviate towards higher $E_{{\rm p}, z}$ than predicted by the Amati relation. 
This occurs because for an observer at an angle $\theta_v$ off the jet axis, the dominant emission comes from regions interior to $\theta_v$. For highly relativistic jets or large viewing angles, relativistic beaming enhances this effect, making the effective emitting region significantly closer to the jet axis than $\theta_v$ itself.
The observed peak energy scales linearly with the Doppler factor, $E_{{\rm p}, z} \propto \mathcal{R}_{\mathcal{D}}$, while the isotropic-equivalent energy scales more steeply, $E_{\gamma, \rm iso} \propto \mathcal{R}_{\mathcal{D}}^3$. 
As a result, the emission appears harder (higher $E_{\rm p}$) but less energetic (lower $E_{\gamma, \rm iso}$) than would be inferred from emission beamed directly along $\theta_v$, upon which the Amati relation was originally formulated. This behavior is qualitatively consistent with analytic models in which the observed
$E_{\rm p}$--$E_{\rm iso}$ relation emerges from the underlying prompt-emission
physics \citep{xu_23}.

In contrast, some FXTs occupy a region $\gtrsim 3\sigma$ \textit{below} the Amati relation. Their spectra are significantly softer than predicted by empirical correlations of Type II GRBs. 
This behavior cannot be explained by viewing-angle effects within the phenomenological model studied here, regardless of the adopted jet structure.
While off-axis emission reduces the observed $E_{\gamma , \rm iso}$, it does not produce the degree of spectral softening required to match the FXT population.
Instead, at least some FXTs likely require additional intrinsic emission variations beyond relativistic and geometric effects.

Studies have constrained the local volumetric event rate of Type II GRBs to be $\sim 1 \rm \, Gpc^{-3} \, yr^{-1}$~\citep{frail_01, wanderman_piran_10}, with the rate increasing by a factor of a few at higher redshifts ($z \gtrsim 3$)~\citep{lan_19, ghirlanda_22}. Assuming a typical beaming factor for Type II GRBs, the intrinsic rate is of order $10^2$--$10^3 \rm \, Gpc^{-3} \, yr^{-1}$. 
In our study, we simulate $7{,}000$ bursts with viewing angles $\theta_v < 30^\circ$. Scaling by the solid angle of the simulated observing cone, this corresponds to $\sim 5\times10^4$ bursts over $2\pi$ steradians. Out of $7{,}000$ bursts with $\theta_v < 30^\circ$, $\sim 250$ are detectable, implying a detection fraction of $\sim 1/200$ that is consistent with observational constraints.

\section{Discussion}
\label{sec:discussion}

\subsection{Comparison with Observations and Previous Work}

As shown in Fig.~\ref{fig:sim_type2} and~\ref{fig:sim_type1}, the most energetic bursts are predominantly viewed on-axis ($\theta_v < \theta_j$; triangles), while observers in the wing region ($\theta_j < \theta_v < \theta_{\rm cut}$; squares) measure bursts with intermediate to low $E_{\rm iso}$. In the multi-component model the transition from the tophat core to the weak power-law wing is sharp, whereas the single-component Gaussian jet shows a smooth exponential decline. Nevertheless, in both cases the wing contributes detectable emission for off-axis observers, allowing bursts with lower $E_{\gamma, \rm iso}$ to remain broadly consistent with the global $E_{\gamma, \rm iso}$--$E_{{\rm p}, z}$ correlation.

Simulated bursts with the multi-component structure begin to overlap the region of XRFs that occupy the low-energy extension of the Amati relation, including events such as GRB 020903~\citep{sakamoto_04} and GRB 060218~\citep{campana_06, pian_06, soderberg_06}.
Detailed studies of XRFs invoke shock breakout from mildly relativistic cocoons \citep[e.g.,][]{wang_07}, baryon-loaded (``dirty'') fireballs with intrinsically low Lorentz factors \citep[e.g.,][]{dermer_99, huang_02}, or prolonged central engine activity \citep[e.g.,][]{lazzati_07_xrf, metzger_11}, rather than viewing angle effects of canonical GRBs. Therefore, this overlap should not be regarded as an explanation for the origin of XRFs, but only as evidence of their consistency with the global Amati relation.

Off-axis emission ($\theta_v > \theta_{\rm cut}$; circles) populates the lowest-energy region of the distribution. This behavior is consistent with models of GRB 170817A as a narrow structured jet viewed off-axis~\citep{mooley_18, troja_18_struct_jet, troja+20_170817, ryan_24, chen_25_xray_grb}, and with Type II llGRBs such as GRB 980425, GRB 031203, and GRB 171205A~\citep{980425, 031203, delia_18}, where the prompt emission may be interpreted as arising from a relativistic jet observed at large viewing angles~\citep{granot_02, yamazaki_03, ramirez_05}. Numerical studies have likewise produced spectrally hard outliers from GRB jets viewed off-axis~\citep[e.g.][]{farinelli_21}. On the other hand, several works suggest that viewing-angle effects alone may not be sufficient to interpret all llGRBs~\citep{delia_18}, and that observational signatures of jet-driven cocoon emission provide strong evidence of additional distinct emission components~\citep{izzo_19, maity_21}.

\subsection{Implications for FXTs and Alternative Interpretations}

EP-detected FXTs without gamma-ray counterparts lie below the $3\sigma$ region of the Amati relation, with $E_{{\rm p}, z}$ below nearby XRFs by $\geq 0.5$ dex.
This offset represents a lower limit on the true deviation, since the true spectral peaks of FXTs likely lie below the EP bandpass.
Larger viewing angles shift the expected $E_{{\rm p}, z}$ toward higher values, opposite of the trend observed among FXTs.

Within our numerical framework, prompt emission properties are tied to empirical correlations between the initial Lorentz factor and isotropic energy ($\Gamma_0$--$E_{\gamma, \rm iso}$)~\citep{liang+10_relation, ghirlanda+11_relation}, together with the Amati relation linking $E_{{\rm p}, z}$ and $E_{\gamma, \rm iso}$~\citep{amati_02}. These correlations combined with viewing-angle effects successfully reproduce the observed distribution of classical GRBs, and populate regions of the $E_{\gamma, \rm iso}$--$E_{{\rm p}, z}$ inhabited by XRFs and llGRBs. However, even when accounting for viewing-angle effects and various outflow structures under the observational prescription, the location of EP-detected FXTs in the $E_{\gamma, \rm iso}$--$E_{{\rm p}, z}$ plane are $> 3\sigma$ outliers. This result suggests that the empirical correlations derived from the observed population of GRBs do not straightforwardly extend to the FXT population.

%One possible explanation is 
Such inconsistency suggests that the physical conditions governing prompt emission in FXTs differ from those in classical GRBs. Variations in prompt emission efficiency, Lorentz-factor structure, or dissipation mechanisms could modify the relationship between $E_{\gamma, \rm iso}$ and $E_{{\rm p}, z}$. In internal dissipation models, the spectral peak depends sensitively on the bulk Lorentz factor, baryon loading, and the microphysics of particle acceleration and magnetic fields~\citep[for comprehensive reviews of GRB physics, see e.g.,][]{kumar_zhang14_grb, zhang18_grb_book}. Changes in internal shock conditions or the efficiency of magnetic reconnection can therefore shift the spectral peak to lower energies without a proportional reduction in the total emitted energy~\citep{mochkovitch_15}. Alternatively, FXTs may arise from outflows with systematically lower Lorentz factors, which would naturally produce softer spectra and weaker Doppler boosting compared to typical GRB jets~\citep{dermer_99, daigne_02}. Indeed, the data of EP241113a \citep{dai_26} are consistent with originating from a ``dirty fireball'' with a Lorentz factor an order of magnitude lower than typical GRBs, a regime that was not observationally verified in the past. 

%BZ: This paragraph may not needed because SBOs have a much lower luminosity/energy than the ones we study.
%Another possibility is that FXTs originate from physically distinct scenarios rather than from standard GRB jets viewed under unusual geometries. Several theoretical models predict soft, X-ray--dominated transients associated with the early stages of relativistic outflows. 
%These include shock breakout emission from a relativistic jet or cocoon emerging from the progenitor star~\citep{waxman_07, nakar_12}, as well as radiation from mildly relativistic outflows produced by failed or choked jets~\citep{bromberg_11, nakar_17}. In such cases, the prompt emission may arise from a different dynamical regime or primary radiation mechanism than those responsible for classical GRBs, potentially producing lower spectral peaks and different energetics. Observationally, this interpretation is supported by the association of some llGRBs and X-ray transients with supernova shock breakout or cocoon emission, such as GRB 060218~\citep{waxman_07} and GRB 100316D~\citep{starling_11}.

%Taken together, these considerations suggest that 
In general, EP-detected FXTs may probe a regime of relativistic outflows that is not well represented in the classical GRB population used to establish empirical prompt-emission correlations. This may be intrinsically related to an intermediate class of progenitor stars between those generating typical GRBs and those generating more mundane supernovae. Compared with GRB progenitor stars, these stars might have a smaller angular momentum or a weaker magnetic field at the core, so that the jet may carry a lower power and entrain more baryons. Detailed studies of these intermediate progenitor stars are desired to better understand EP-detected FXTs. 
%Continued observations with wide-field soft X-ray instruments will therefore be essential for determining whether FXTs represent the low-energy extension of GRB phenomenology or a distinct class of high-energy transients.

\subsection{Caveats}
The primary aim of our simulations is to investigate the observational signatures of different jet structures, rather than to reproduce precisely the observed sample \citep[e.g.][]{mp_20}. 
A direct comparison with observational datasets is complicated by strong selection effects: GRB samples are shaped by detector sensitivity, energy band coverage, and viewing-angle-dependent detectability. Instruments preferentially detect bursts with higher fluence within their bandpass, which biases observational samples.
Observational samples are further affected by incomplete redshift measurements and host-galaxy extinction, meaning that the detected population does not uniformly sample the intrinsic GRB distribution. 
As a result, precise fits to observational data are not a reliable benchmark for theoretical models, though volumetric event rates can still provide qualitative guidance.

The intrinsic angular structure of GRB jets, whether single-component or multi-component, is not directly observable and remains uncertain. 
The Amati relation, $E_{\rm p} \propto E_{\rm iso}^a$ (Eq.~\ref{eq:amati}), is a phenomenological correlation derived from the observed population, which likely includes both on-axis and off-axis events, as well as contributions from jet cores and wings or cocoons. 
Applying this relation to off-axis emission is therefore approximate, assuming the physics governing the prompt emission is broadly similar across viewing angles and jet components.

Additional uncertainties arise from poorly constrained physical parameters, such as circumburst density, radiative efficiency, and microphysical shock parameters. These introduce scatter beyond that captured by parameterized outflow models. As we do not consider intrinsic microphysical parameters nor particular radiation mechanisms, our simulations are intended to provide a first-order approximation of population-level trends and qualitative signatures of different jet structures, rather than to reproduce individual observed bursts or exact distributions.

A caveat of our detectability implementation is that we model light curves as smooth FRED pulses, which lack the rapid variability commonly observed in GRB light curves. Such variability can significantly enhance peak flux on short timescales, increasing the probability of triggering. As a result, our approach provides a conservative estimate of detectability.

\section{Conclusion}
\label{sec:conclusion}

In this work, we apply the numerical framework developed in \cite{chen_25_xray_grb} to investigate the growing sample of FXTs detected by EP. We consider two physically motivated and structurally distinct relativistic outflow structures relevant to Type II GRB progenitors: a single-component Gaussian and a multi-component jet consisting of a narrow core surrounded by a broader wing with lower energy and Lorentz factor. Using simulated populations of GRBs that incorporate empirical prompt-emission correlations, viewing-angle effects, and detectability, we explore how these outflow structures populate the $E_{\gamma, \rm iso}$--$E_{{\rm p}, z}$ plane.

Our results show that single-component jets viewed off-axis tend to exhibit elevated $E_{\rm p}$ relative to $E_{\gamma, \rm iso}$, consistent with previous studies of structured jets and off-axis emission~\citep{granot_02, yamazaki_03, ramirez_05, dado_17, farinelli_21, ryan_24, chen_25_xray_grb}. In these models, the combined effects of relativistic Doppler boosting and angular energy structure cause the observed spectral peak to remain relatively high even as the observed isotropic-equivalent energy decreases with viewing angle. This deviation is opposite of the trend observed in FXTs, thus we conclude that viewing angle effects of canonical Type II GRBs alone do not reproduce energetics observed among FXTs.

The multi-component jet structure considered here includes a broader wing with lower energy and Lorentz factor, which dominates the observed emission outside the narrow core. While this extended structure pushes the population of detectable bursts toward lower $E_{\rm p}$ and $E_{\gamma, \rm iso}$, we find that it still does not fully populate the distinct parameter space occupied by EP-detected FXTs. We note that this lower-energy regime is broadly consistent with XRFs~\citep{sakamoto_04, irwin_16}, though this does not provide conclusive evidence for their physical origins. Furthermore, ``true'' off-axis observers beyond both the jet and wing components ($\theta_v > \theta_{\rm cut}$) measure spectrally hard bursts relative to $E_{\gamma, \rm iso}$, consistent with the energetics of llGRBs~\citep{granot_02, yamazaki_03, ramirez_05}. 

Our results suggest that FXTs probe physical regimes not captured by standard phenomenological frameworks used to describe canonical Type II GRBs. Several possibilities may account for this discrepancy. The prompt emission physics may differ from that of typical GRB jets, for example through variations in Lorentz-factor structure, baryon loading, or radiation mechanisms that modify a phenomenological relationship between $E_{\gamma, \rm iso}$ and $E_{{\rm p}, z}$. %Alternatively, FXTs may arise from physically distinct progenitors such as shock breakout emission, cocoon-dominated outflows, or mildly relativistic jets produced by failed or choked central engines~\citep{waxman_07, bromberg_11, nakar_17}. 

The rapidly increasing FXT sample discovered by EP provides a new opportunity to explore this previously under-sampled region of parameter space. By extending prompt transient searches into the soft X-ray regime with wide-field monitoring, EP is revealing events that were largely inaccessible to gamma-ray instruments. Determining whether FXTs represent the low-energy extension of the GRB population or a distinct class of relativistic transients will require continued multiwavelength follow-up, improved spectral constraints, and population studies that incorporate a broader range of outflow structures and radiation mechanisms.

Future work will focus on extending the present framework to include additional physical ingredients, such as refined angular energy and Lorentz-factor distributions from numerical simulations based on first principles, direct modeling of intrinsic prompt emission mechanisms, and alternate progenitor scenarios beyond a typical collapsar jet. Incorporating detailed radiation physics and confronting these models with the expanding EP dataset will be essential for understanding the origin of FXTs and, more broadly, for clarifying the diversity of relativistic explosions produced during stellar collapse.

\section*{Acknowledgements}
The authors acknowledge the Nevada Center for Astrophysics and a UNLV Top-Tier Doctoral Graduate Research Assistantship (TTDGRA) for support. C.C acknowledges Dr. Mike Moss (NASA GSFC) and Dr. Brad Cenko (NASA GSFC) for helpful discussions and feedback.

\appendix
\section{Konus-Wind Detectability}
\label{app:kw_detectability}

\begin{figure*}[h!]
    \centering
    \includegraphics[width=\linewidth]{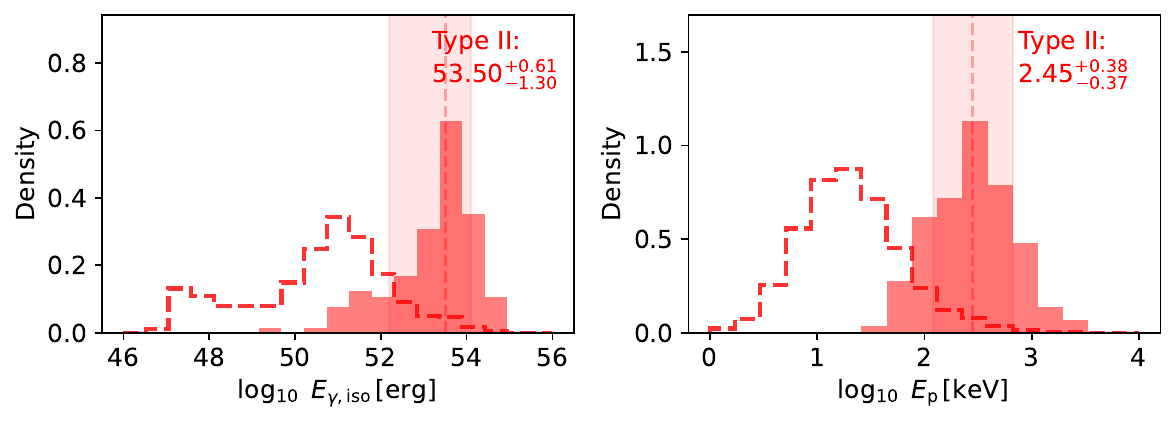}
    \caption{$E_{\gamma, \rm iso}$ and $E_{\rm p}$ distribution densities for Type II GRBs with a multi-component jet structure, assuming KW detectability. KW-detectable bursts are represented by the filled histogram, while the full simulated sample including undetectable bursts is shown as the dashed step histogram. Compared to Swift/BAT-detectable bursts (Fig.~\ref{fig:multi_dist}), KW-detectable bursts have higher $E_{\gamma, \rm iso}$ and $E_{\rm p}$ due to the higher energy bandpass of KW.}
    \label{fig:multi_dist_KW}
\end{figure*}

\begin{figure}[h!]
    \centering
    \includegraphics[width=.49\linewidth]{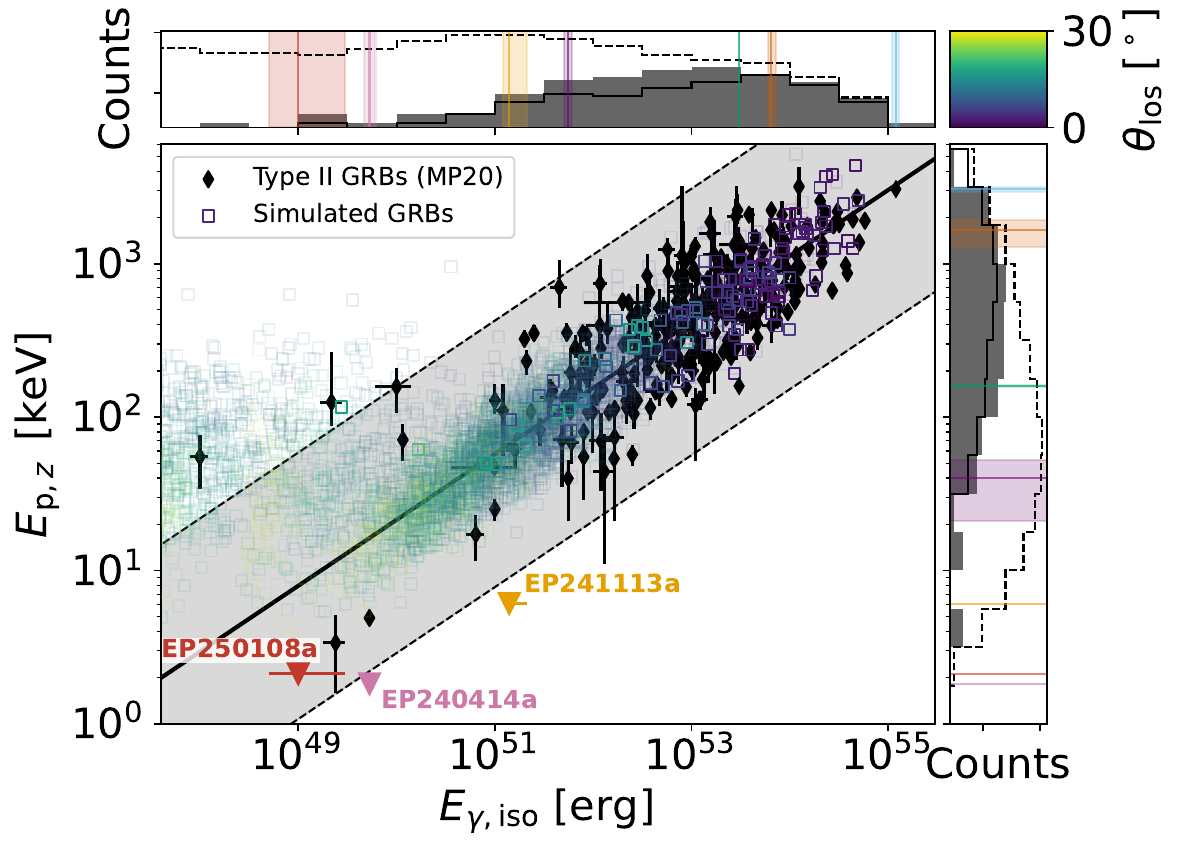}
    \caption{
    Simulated and observed Type II GRBs with the multi-component jet structure on the $E_{\gamma, \rm iso}$--$E_{{\rm p}, z}$ plane.
    Opacity indicates whether a burst is detectable by KW (opaque) or not (semi-transparent). For comparison, the Swift/BAT-detectable population is shown in the right panel of Fig.~\ref{fig:sim_type2}.
    }
    \label{fig:sim_KW}
\end{figure}

A significant fraction of Type II GRBs in the observational sample are detected by Konus-Wind (KW), which operates in a higher energy bandpass ($80$--$1{,}200$ keV) than Swift/BAT ($15$--$150$ keV). 
As a consistency check, we evaluate detectability using an approximate KW sensitivity threshold of 
\begin{equation}
f_{{\rm KW}, 9\sigma} = 10^{-6} \left(\frac{T}{1\,\rm s}\right)^{-0.5} \rm \, erg\,cm^{-2}\,s^{-1}
\end{equation}
for two fixed triggering timescales, $T = 140 \rm \, ms$ and $T = 1 \rm\, s$~\citep{kw_17}. 
The resulting distribution densities of $E_{\gamma, \rm iso}$ and $E_{\rm p}$ for KW-detectable bursts, compared to the full simulated sample, are shown in Fig.~\ref{fig:multi_dist_KW}. As expected, the higher-energy bandpass preferentially selects bursts with larger $E_{\rm p}$ and $E_{\gamma, \rm iso}$, while many softer bursts ($E_{\rm p} \leq 70 \rm \, keV$) fall below the detection threshold.
The population of simulated and observed Type II GRBs with the multi-component jet structure on the $E_{\gamma, \rm iso}$--$E_{{\rm p}, z}$ plane is shown in Fig.~\ref{fig:sim_KW}. 
Given their soft spectra and low $E_{\rm p}$, FXTs typically fall below the KW sensitivity threshold and would not be expected to trigger the instrument.

\clearpage

\bibliography{main}{}
\bibliographystyle{aasjournalv7}

\end{document}